\documentstyle[aaspp4,12pt,psfig]{article}
\lefthead{Bower et al.}
\righthead{Linear Polarization of Sgr~A*}
\begin{document}

\newcommand\degd{\ifmmode^{\circ}\!\!\!.\,\else$^{\circ}\!\!\!.\,$\fi}
\newcommand{\etal}{{\it et al.\ }}
\newcommand{\uv}{(u,v)}
\newcommand{\rdm}{{\rm\ rad\ m^{-2}}}

\title{The Linear Polarization of Sagittarius A* I.  VLA Spectro-polarimetry at
4.8 and 8.4 GHz}
\author{Geoffrey C. Bower\altaffilmark{1,2}, 
Donald C. Backer\altaffilmark{3}, 
Jun-Hui Zhao\altaffilmark{4}, 
Miller Goss\altaffilmark{2} \&
Heino Falcke\altaffilmark{1}}
\altaffiltext{1}{Max Planck Institut f\"{u}r Radioastronomie, Auf dem H\"{u}gel 69, D 53121 Bonn Germany}
\altaffiltext{2}{National Radio Astronomy Observatory, P.O. Box O, 1003 Lopezville, Socorro, NM 87801}
\altaffiltext{3}{Astronomy Department \& Radio Astronomy Laboratory, University of California, Berkeley, CA 94720}
\altaffiltext{4}{Harvard-Smithsonian Center for Astrophysics, Mail Stop 42, 60 Garden Street, Cambridge, MA 02138}

\centerline{\it Accepted for publication in the Astrophysical Journal}

\begin{abstract}

Synchrotron radiation from active galactic nuclei (AGN)
is often highly polarized.
We present a search for linear polarization with the Very Large Array (VLA)
at 4.8 GHz and 8.4 GHz from the nearest AGN, Sagittarius A*. 
As a part of this study we used  spectro-polarimetric data 
that were sensitive to a rotation measure (RM) as large as 
$3.5\times 10^6\rdm$ at
4.8 GHz and $1.5\times 10^7\rdm$ at 8.4 GHz.  The upper limit to the
linear polarization of Sgr~A* over a broad range of RM is 0.2\% at
both frequencies.  We also present continuum observations with the 
VLA at 4.8 GHz which give an upper limit of 0.1\% for RMs less than
$10^4 \rdm$.  We conclude that depolarization is unlikely to occur
in the Galacter Center scattering medium.
However, it is possible for depolarization to occur in the accretion
region of Sgr A* if the outer scale of turbulence is small enough.
We also consider the implications of a very low
intrinsic polarization for Sgr~A*.

\end{abstract}


\section{Introduction}

The compact non-thermal radio source Sgr~A* has long been recognized as
a massive black hole candidate.   Recent results from
stellar proper motion studies indicate that there is a 
dark mass of $\sim 2.6 \times 10^6 M_{\sun}$ enclosed within 0.01 pc
(Genzel \etal \markcite{genze97} 1997, Ghez \etal \markcite{ghez98} 1998).
Very long baseline interferometry studies at millimeter wavelengths
have shown that the intrinsic radio source coincident with the dark mass
has a size that is less than
1 AU and a brightness temperature greater than $10^9$ K (Rogers \etal 
\markcite{roger94} 1994,
Bower \& Backer \markcite{bower98} 1998, Lo \etal \markcite{lo98} 1998).  Together these points are compelling
evidence that Sgr~A* is a cyclo-synchrotron emitting region surrounding
a massive black hole.  Nevertheless, specific details of the excitation of high
energy electrons, their distribution and the accretion of infalling
matter onto Sgr~A* are unknown (e.g., Falcke, Mannheim \& Biermann \markcite{falck93} 1993, 
Melia \markcite{melia94} 1994, Narayan \etal \markcite{naray98} 1998, Mahadevan \markcite{mahad98} 1998).

Linear polarization stands as one of the few observables of Sgr~A*
not extensively investigated observationally or theoretically.  However,
we expect linear polarization to arise from the cyclo-synchrotron
radiation that is responsible for the radio to millimeter wavelength
spectrum.  A homogeneous, optically-thin, synchrotron 
source with a uniform magnetic
field has a fractional polarization of 70\%.   Measured fractional
polarizations in AGN are typically a few percent at 
wavelengths shorter than 6 cm where the compact cores dominate
the highly-polarized radio lobes in 
the total flux (e.g., Aller, Aller
\& Hughes \markcite{aller92} 1992).
However, polarization VLBI images sometimes show regions
of significantly enhanced polarization 
(Brown, Roberts \& Wardle \markcite{brown94} 1994).  

The polarization of Sgr~A* may prove to be as important a diagnostic of
models for the radio to millimeter spectrum as it has been for AGN.  
Detection of linear polarization in AGN has firmly established 
synchrotron emission as the radiation mechanism.
Comparison of the evolution
of linear polarization to the evolution of total intensity has provided
a strong argument for the existence of shocks in the relativistic
jets of AGN (e.g., Hughes, Aller and Aller \markcite{hughe85} 1985).  
Detection of similar 
correlations in polarized and total intensity variations in Sgr~A* would be
convincing evidence for a jet.  Other models may have unique
signatures for polarized intensity variations.

Sgr~A* is located in a region with strong magnetic fields and high
electron density.  The image of Sgr~A* is significantly scatter-broadened
by intervening thermal plasma (e.g., Lo \etal \markcite{lo98} 1998), as are the images
of many masers in the Galactic Center region (Frail \etal \markcite{frail94} 1994).
Furthermore, nonthermal filaments in the Galactic
Center region show RMs which vary on the arcsecond scale and
are as large as 4000 $\rdm$ (Yusef-Zadeh, Wardle \& Parastaran \markcite{yusef97} 1997).
Such large RMs can effectively depolarize a signal detected with a
large bandwidth.

In \S 2, we discuss the effect of large RMs on a polarized signal
and our Fourier transform technique for detecting large RMs.
In \S 3.1, we present VLA
continuum observations at 4.8 GHz.
In \S 3.2 and \S 3.3, we present VLA spectro-polarimetric observations 
at 4.8
GHz and 8.4 GHz.  These observations are sensitive to a wide-range of 
RMs.  In \S 4, we consider other effects of 
interstellar matter on a polarized signal from Sgr~A*.  And in
\S 5, we discuss the consequences of our upper limits for the
polarization on models for Sgr~A*.  In a future paper, we will address
millimeter polarization observations of Sgr~A*.

\section{Searching for Large RMs}

In an ionized and magnetized region
right and left circularly polarized waves will have different
indices of refraction.  This leads to a wavelength-dependent
delay between circular polarizations which 
is equivalent to a rotation of the position angle $\chi$ of a 
linearly polarized signal
\begin{equation}
\chi_F = {\rm RM}\ \lambda^2,
\end{equation}
where RM is the rotation measure.
This rotation of $\chi$ is equivalent to a rotation in the two-dimensional
Stokes
$Q$ and $U$ space.

A linearly polarized signal will be significantly depolarized in
an observing bandwidth $\Delta\nu$ if $\chi$ rotates by more than one
radian, or if the RM exceeds
\begin{equation}
{\rm RM_{max}}={1\over 2 } {1 \over \lambda^2}{\nu \over \Delta\nu}.
\label{eqn:rmmax}
\end{equation}
If the bandwidth $\Delta\nu$ is split into $n \times \delta\nu$ channels,
a search can be made for RMs larger than ${\rm RM_{max}}$.
When $\chi$ wraps through more than
one turn, $n\pi$ ambiguities make it impossible 
through a linear least squares fit to detect RMs.
Fourier transforming
the complex visibility $P=Q + i U$ with respect to $\lambda^2$
searches for large RMs without loss of sensitivity.  
The maximum RM detectable in
this scheme can be found by replacing $\Delta\nu$ with $\delta\nu$ in
Equation~\ref{eqn:rmmax}.  
In addition to detecting RMs that exceed ${\rm RM_{max}}$, the technique
is sensitive to multiple RMs from the same object.  A more detailed
analysis of this technique can be found in Killeen \etal \markcite{kille99} (1999).

A continuum observation with the VLA at 4.8 GHz with 50 MHz has 
${\rm RM_{max}} \approx 10^4 \rdm$.  
Splitting the band into 256 channels increases 
${\rm RM_{max}}$ by two orders of magnitude to
$3.5\times 10^6 \rdm$.  
The minimum fully-sampled RM detectable in a spectro-polarimetric data set,
${\rm RM_{min}}$, is 
approximately equal to ${\rm RM_{max}}$ for a continuum data set with
the same total bandwidth.

The RM can be found to better accuracy than ${\rm RM_{min}}$.  We estimate
the error to be 
\begin{equation}
\sigma_{\rm RM} = { {\rm RM_{min} } \over {\rm SNR} } .
\end{equation}
SNR is the ratio of the peak amplitude in Fourier space to the off-peak
root-mean-square noise.

\section{Observations and Data Reduction}

\subsection{VLA Continuum Polarimetry at 4.8 GHz}

The VLA of the National Radio Astronomy Observatory\footnote{The 
National Radio Astronomy Observatory is a facility 
of the National Science Foundation operated under cooperative agreement 
by Associated Universities, Inc.}
observed Sgr~A* on 10 and 18 April 1998 in the A array at 4.8 GHz
with a bandwidth of 50 MHz.  Instrumental calibration was performed
with the compact sources 1741-038 and 1748-253.  The right-left phase
difference was set with observations of 3C~286.  Only baselines longer
than 100 $k\lambda$ were used for Sgr~A*.  Several nearby calibrator
sources, GC 441, W56 and W109, were also observed (Backer \& Sramek \markcite{backe99} 1999).
All sources were self-calibrated and imaged in Stokes $I, Q$ and $U$.

We summarize the measured polarized and total intensities
of Sgr~A* and the calibrators in Table~\ref{tab:vla6cm}.  
The rms noise in the Sgr~A* map is 74 $\ \mu {\rm Jy}$.
Consistency between the results on the two days indicates the
accuracy of the results.   Polarization was reliably detected from
all sources but Sgr~A* and GC 441.  The measured polarization
at the position of
Sgr~A* is 0.1\%.  This value is equal to the average off source
fractional polarization in the map and is, therefore, an upper limit.
The maximum RM detectable with this bandwidth is $\sim 10^4 \rdm$.

\subsection{VLA Spectro-polarimetry at 4.8 GHz}

The VLA observed Sgr~A* in the A array in a spectro-polarimetric mode
at 4.8 GHz on 27 November 1992.   Observations were carried out in
8 consecutively-spaced frequency bands of 6.25 MHz each.  Each band
was divided into 32 separate frequency channels.  The bands
covered the frequency range from 4832 MHz to 4882 MHz.  Five scans
of 2.5 minutes apiece on Sgr~A* were interleaved with six scans of
2.5 minutes apiece on NRAO~530 in each frequency band.  Amplitude,
phase and polarization calibration were performed separately for each
band.  Polarization calibration was performed with NRAO~530 alone
and with NRAO~530 and Sgr~A*, producing similar final results.
The right-left phase difference was set for each band with
an observation of 3C~286.

\begin{figure}
\plotone{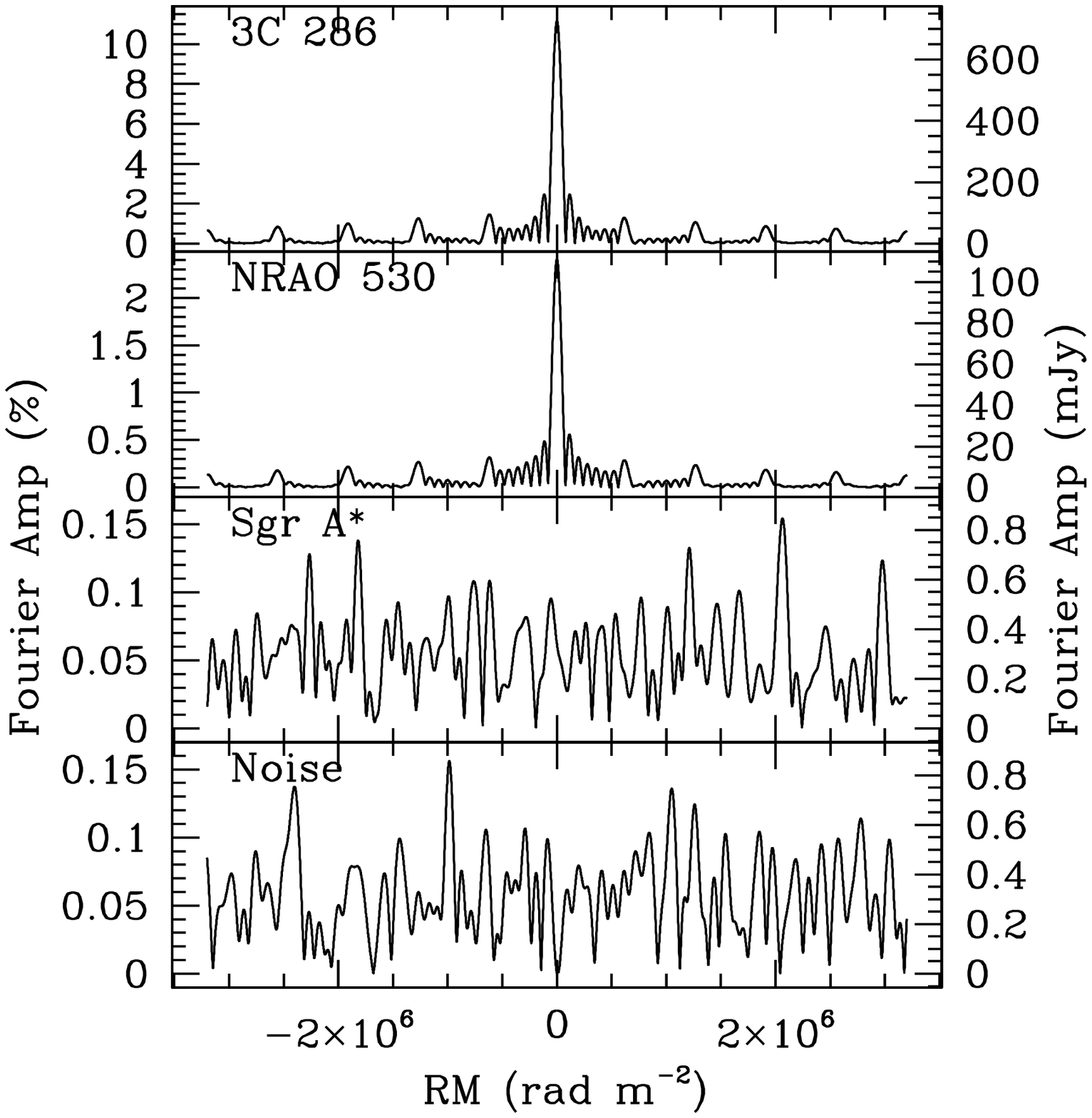}
\caption{The Fourier amplitude for 3C 286, NRAO 530, Sgr A*
and a noise data set at 4.8 GHz.  The Fourier amplitude is given in mJy and
as a
fraction of the total flux for each source.  
The scaling of the Gaussian noise data is set to match that of Sgr A*.
The RM is plotted from $-3.5\times 10^6 \rdm$ to $3.5 \times 10^6 \rdm$.
\label{fig:fouramp6cm}} 
\end{figure}

For each source, the spectral data were
time-averaged and exported from AIPS for further processing.
A bandpass correction was applied.  The complex polarization was
then Fourier-transformed with respect to $\lambda^2$.  Sampling
effects were removed through a one-dimensional CLEAN method.
The CLEAN method permits a better estimate of the RM peak
and of the noise level.  The sampling sidelobes are readily
visible for 3C~286 and NRAO~530 in Figure~\ref{fig:fouramp6cm}.
Our tests with noise data and with synthetic signals indicate 
that the CLEAN method does not generate false signals and
improves the accuracy of peak determination.  Applying CLEAN
to the 4.8 GHz NRAO~530 data reduced the noise in the spectrum
from 1.6 mJy to 0.26 mJy.

The range of fully-sampled RM is $10^4 \rdm$
to $3.5\times 10^6 \rdm$.  The Fourier amplitude for each 
source is shown in Figure~\ref{fig:fouramp6cm} and the results
are summarized in Table~\ref{tab:rm6cm}.  These images are without
bandpass correction and dirty-beam removal.  We also calculate and
plot the Fourier transform for a distribution of Gaussian noise.
Strong peaks at low RM are apparent
for both 3C~286 and NRAO~530,
as expected.  The measured  values are
consistent with the known RMs of these sources:  $1\pm 2 \rdm$ for
3C~286 and $-63 \pm 5 \rdm$ for NRAO~530 (Rusk \markcite{rusk88} 1988).

No strong peak is apparent for Sgr~A* at any RM.  The maximum Fourier
amplitude for Sgr A* is $0.15\%$ at ${\rm RM}=2.1 \times 10^6 \rdm$.
Imaging Sgr A* with and without a RM correction produced a peak polarization
of $0.2\%$.  This is equal to the fractional polarization of 
thermal ionized gas in the vicinity of Sgr A*, indicating that we are 
limited by residual instrumental polarization.

\subsection{VLA Spectro-polarimetry at 8.4 GHz}

The VLA observed Sgr~A* in the A array in a spectro-polarimetric mode
at 8.4 GHz, also on 27 November 1992.   Observations were carried out in
7 frequency bands of 6.25 MHz each.  Each band
was divided into 32 separate frequency channels.  Five bands
covered the frequency range from 8405 MHz to 8437 MHz.  Two other
bands were centered at 8150 MHz and 8700 MHz.  Five scans
of 2.5 minutes apiece on Sgr~A* were interleaved with six scans of
2.5 minutes apiece on NRAO~530 in each frequency band.  
Amplitude,
phase and polarization calibration were performed separately for each
band.  Polarization calibration was performed with NRAO~530 alone
and with NRAO~530 and Sgr~A* together, producing similar final results.
The right-left phase difference was set for each band with
an observation of 3C~286.  
The sources W56, 1741-312, GC 441, W109 
and 1748-253 were observed for two minutes in the three 6.25 MHz bands
centered at 8150 MHz, 8420 MHz and 8700 MHz.
The results for all sources were the same using all frequency bands or
only the inner 5 bands.
The same reduction steps were taken for the 8.4 GHz data as for the 4.8
GHz data.

\begin{figure}
\plotone{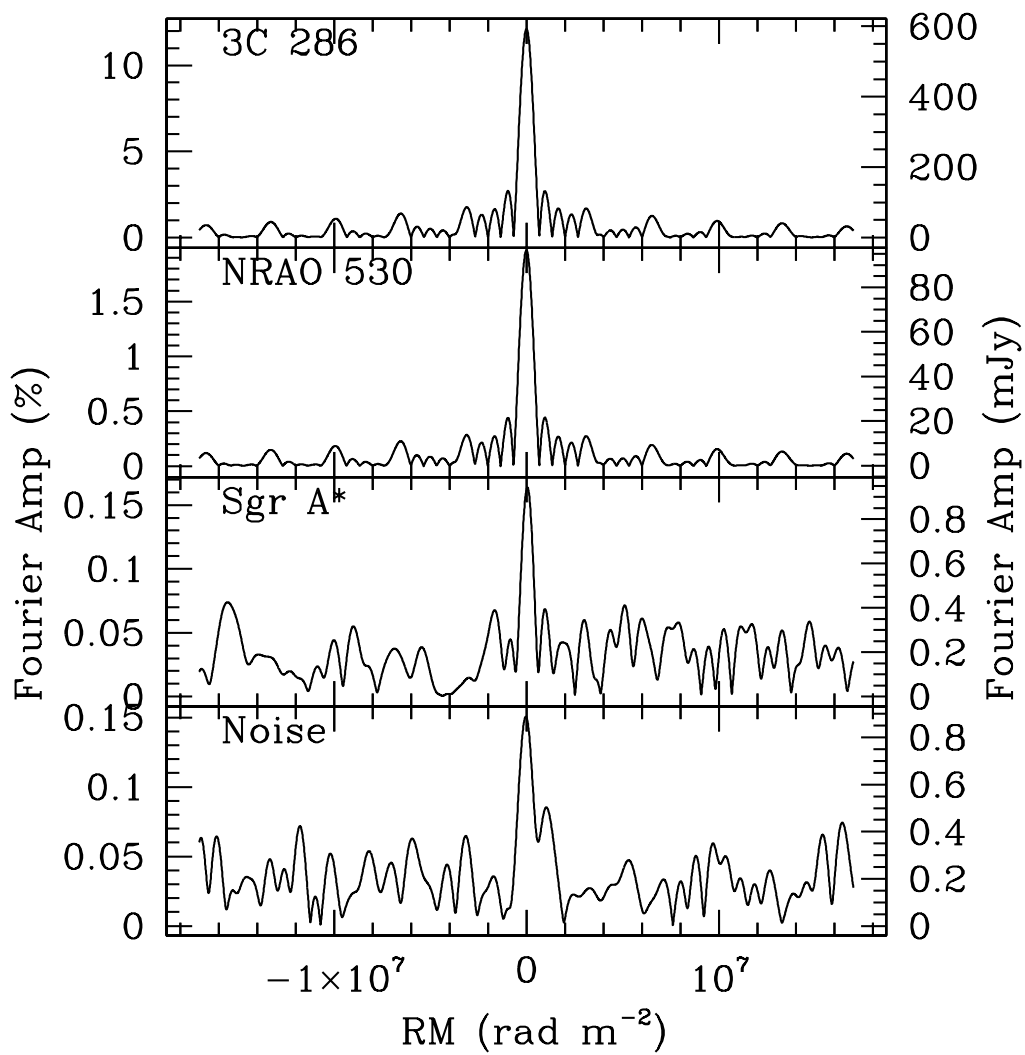}
\caption{The Fourier amplitude for 3C 286, NRAO 530, Sgr A*
and a noise data set at 8.4 GHz.  The Fourier amplitude is given in mJy and
as a
fraction of the total flux for each source.  
The noise data consists of Gaussian noise scaled to the level of Sgr A*
and an 0.15\% fractional polarization at zero RM.
The RM is plotted from $-1.5\times 10^7 \rdm$ to $1.5 \times 10^7 \rdm$.
\label{fig:fouramp4cm1}} 
\end{figure}

\begin{figure}
\plotone{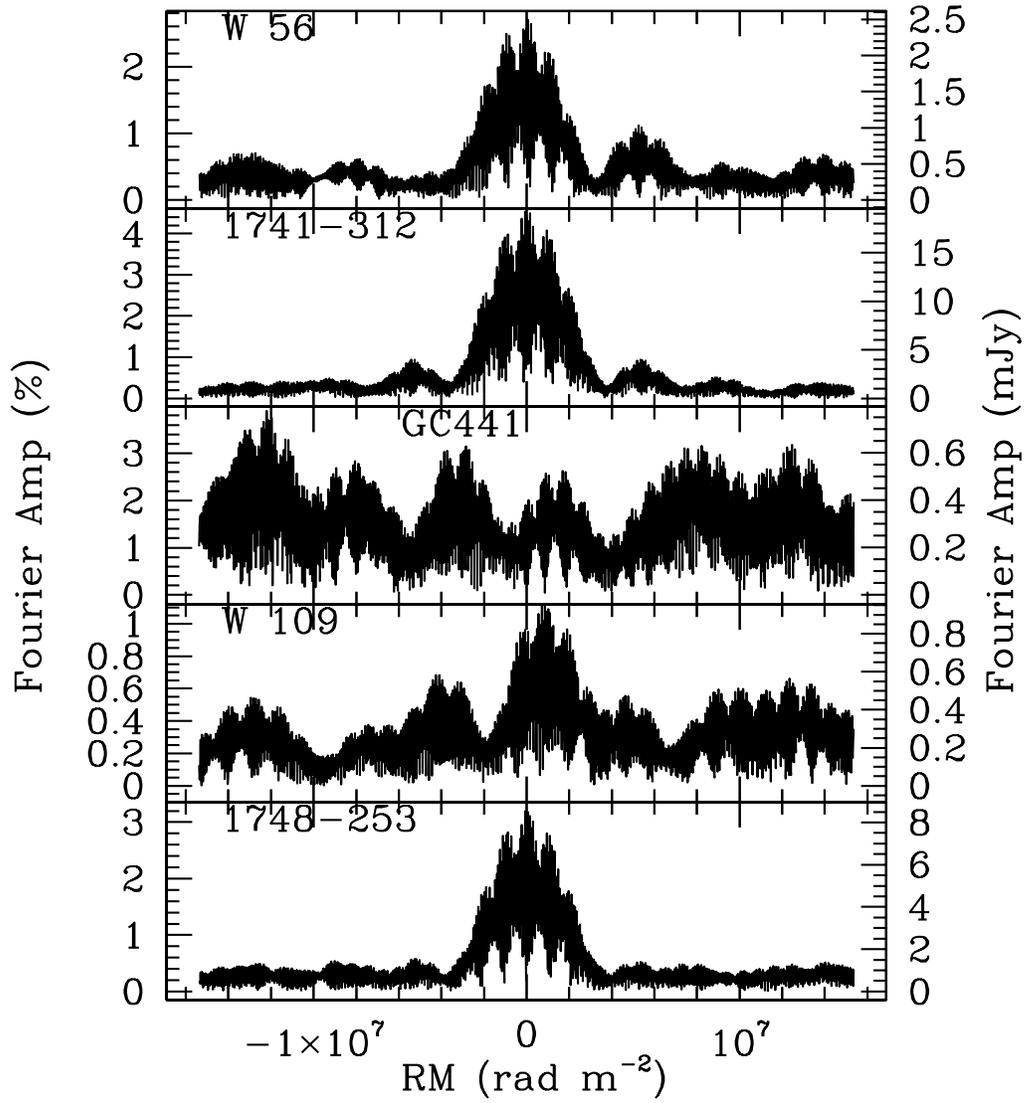}
\caption{The Fourier amplitude for W56, 1741-038, GC441,
W109 and 1748-253 at 8.4 GHz.  The Fourier amplitude is given in mJy and
as a
fraction of the total flux for each source.  
The RM is plotted from $-1.5\times 10^7 \rdm$ to $1.5 \times 10^7 \rdm$.
\label{fig:fouramp4cm2}  }
\end{figure}

The Fourier amplitudes for all sources are shown in 
Figures~\ref{fig:fouramp4cm1} and \ref{fig:fouramp4cm2}. 
These images are without
bandpass correction and dirty-beam removal.   
The results are summarized in Table~\ref{tab:rm3cm}.
These data are sensitive to $3.5 \times 10^5 < | {\rm RM} | < 1.5 \times10^7 \rdm$.
There are
strong detections of linear polarization in 3C~286 and NRAO~530 at
RMs consistent with zero.
Significant detections were also made for W56, W109, 1741-312 and 1748-253,
also at RMs consistent with zero.  No polarization was detected in GC 441.
The errors in RM for these secondary calibrators are larger due to the
sparser frequency coverage and shorter observing time.

For Sgr A*, we detect a peak in the Fourier spectrum of 0.17\% at
${\rm RM}=24000 \pm 37000 \rdm$.  
Imaging Sgr A* with and without RM corrections, 
we find a fractional polarization of 0.1\%.  Off-source
fractional polarizations are typically 0.1\%, again implying that
we are limited by residual instrumental polarization.

We tested noise models to see if we could reproduce a weak signal at non-zero
RM.  We used an input signal with ${\rm RM} = 0$ at 0.15\% of the peak
intensity of Sgr A* and noise that matched that of Sgr A*.  This is
the model plotted in Figure~\ref{fig:fouramp4cm1}.  The 
measured RM peak wandered within the error range.  

\section{Interstellar Propagation Effects}

The interstellar medium may depolarize a linearly
polarized radio wave in two ways:
significant rotation of the polarization position angle through the observing
bandwidth; and, 
differential Faraday rotation along the many paths that contribute to
the scatter-broadened image of Sgr A*. 
We have already addressed the first effect in \S 2 and found in
\S 3 that Sgr A* is not depolarized by RMs less than $1.5\times 10^7 \rdm$.
We now consider the second effect.

The scattering region will depolarize the signal
if $\delta\chi_F\approx\pi$.  For our observing wavelengths,
$\delta{\rm RM}=900 {\rm\ rad\ m^{-2}}$ and
$\delta{\rm RM}=2400 {\rm\ rad\ m^{-2}}$.
Over the scattering size of 50 mas at 4.8 GHz,
this corresponds to $\delta{\rm RM}/\delta{\theta}=18000 {\rm\ rad\ m^{-2}
arcsec^{-1}}$.  

Observed variations in RM in the GC region are many orders of
magnitude below those necessary to depolarize Sgr A*.
Observations on the arcsecond to arcminute 
scale of a nonthermal filament within 1 degree of Sgr A* find a maximum
$\delta{\rm RM}/\delta{\theta}=250 {\rm\ rad\ m^{-2}\ arcsec^{-1}}$
(Yusef-Zadeh, Wardle \& Parastaran \markcite{yusef97} 1997).  
Extrapolation of the 
RM structure function to the scattering size implies
$\delta{\rm RM}\approx 50 {\rm\ rad\ m^{-2}}$.
However, these observations are made on a much larger scale than the
scattering disk of Sgr A* and the scattering medium is believed to be
inhomogeneous.

Could the more extreme conditions necessary
to depolarize Sgr A* exist in the Galactic Center
scattering region?
The RM is expressed as
\begin{equation}
{\rm RM} = 0.8 n_e B L {\rm\ rad\ m^{-2}},
\end{equation}
where $n_e$ is the electron number density in cm$^{-3}$, 
$B$ is the magnetic field parallel to the line of sight
in $\mu{\rm G}$ and $L$ is the size scale in pc.  
Since $L$ must be a fraction of the scattering diameter, we find
$L \sim 0.1 \theta_{Sgr A*} D_{Sgr A*} \sim 10^{-4} {\rm\ pc}$.
Yusef-Zadeh \etal \markcite{yusef94} (1994) argued that the 
photo-ionized skins of molecular
clouds in the GC region have a similar length scale, milliGauss fields 
and $n_e \sim 10^4 {\rm cm^{-3}}$.
This matches the depolarization condition if the regions are fully
turbulent.
However, if the constraints on the outer scale of turbulence derived
by Lazio \& Cordes \markcite{lazio98} (1998) are correct, then $L \sim 10^{-7}{\rm\ pc}$.
In this case, 
the RM condition and pressure balance 
between the magnetic and thermal components can only be satisfied 
if $B\sim 10 {\rm\ mG}$ and $n_e \sim 10^6 {\rm\ cm^{-3}}$.
These conditions are extreme even for the GC region.  The largest
magnetic fields as measured for OH masers 
are on the order of a few milliGauss (e.g., Yusef-Zadeh
\etal \markcite{yusef96} 1996).  Ionized densities measured for H II regions 
on the arcsecond scale ($\la 0.1$ pc) are significantly
less than $10^5 {\rm\ cm^{-3}}$
(Mehringer \etal \markcite{mehri93} 1993).  
No depolarization is predicted for the higher temperature and lower density 
model of Lazio \& Cordes \markcite{lazio98} (1998)  for $B < 1 {\rm\ G}$.  
The conditions necessary to depolarize at 8.4 GHz are even more extreme.
We conclude, therefore, that the conditions necessary to depolarize
Sgr A* are unlikely to occur in the scattering region.

We consider now whether depolarization may occur in
the accretion region of Sgr A*, where the electron density and
magnetic field strength are large but the length scale is smaller.
If we consider the simplest model of spherical infall with 
$\dot{M}=10^{-4} M_{\sun} y^{-1}$ and equipartition between particle,
magnetic and gravitational energy (Melia \markcite{melia94} 1994), 
then the change $\delta{\rm RM}$ over an interval $\delta r$ at a 
radius $r$ from Sgr A* is
\begin{equation}
\delta {\rm RM} = 1.2 \times 10^{14} r^{-11/4} \delta r \rdm,
\end{equation}
where we have expressed $r$ and 
$\delta r$ in units of the gravitational radius
$r_g =2 G M/c^2=7.8\times 10^{11} {\rm\ cm}$ for a $2.6 \times 10^6 M_{\sun}$
black hole.  
This relation only holds for $r \ga 10^{3}$, where the temperature
falls below $10^9 {\rm\ K}$.  The RM inside of this radius is
negligible unless there is a separate population of cold
electrons.  We consider the effect of cold electrons in more detail in
the following Section.
If the scattering screen is at a distance of 100 pc from Sgr A*, then
the image will be an average of ray paths over a tangential length scale
$l \sim 2\times 10^{-5} r$.  
Fluctuations in the RM will depolarize Sgr A*
at a given radius if $\delta{\rm RM} > 900 \rdm$ and $l > l_0$,
where $l_0$ is the outer scale of turbulence.  Assuming
$\delta r \sim r$, we find that depolarization will occur only if
$l_0 \la 10^{-5} {\rm\ pc}$.  Although this scale is much smaller 
than the outer scale in the local ISM (Armstrong, Rickett \& Spangler
\markcite{armst95} 1995), it is a scale that may be pertinent to the dense, energetic
environment of the accretion region.

\section{An Intrinsically Weakly Polarized Sgr A*}

A polarization fraction less than 1\% is uncommon in compact radio
sources at wavelengths shorter than 6 cm 
(Aller, Aller \& Hughes \markcite{aller92} 1992).  However,
optically thick
quasar cores observed with VLBI are frequently weakly polarized
(Cawthorne et al. \markcite{cawth93} 1993).  Such cores may be analogous to the radio source
in Sgr A*, which, due to its low power, may not produce the strong shocks 
in the jet
that are the source for higher polarization regions in quasars.
Weak polarization is
more common in radio galaxies than quasars or blazars and it is also more
common in compact-double 
sources or sources with irregular morphologies (Aller, Aller \& Hughes \markcite{aller92} 1992).
A notable source with a very low polarization fraction ($<0.1\%$)
is the radio galaxy 
3C 84, which has a very irregular morphology.

If the radio to
millimeter spectrum of Sgr A* does arise in a jet, the low power of this jet 
or environmental effects in the Galactic Center region  
may limit the magnetic field order.
For the case of a spherically symmetric emitting region, an ordered
magnetic field may depolarize the source, as well.
Alternatively, low energy electrons in the synchrotron 
environment of Sgr A* may Faraday depolarize the source.
The ADAF model and the Bondi-Hoyle accretion model predict the presence
of non-relativistic electrons in the accretion region.

Observations at millimeter wavelengths may resolve many of
the questions raised in this paper.  Interstellar effects are
reduced such that depolarization in the scattering region is 
extremely unlikely and depolarization in the accretion region must
occur at radii less than 0.01 pc.  Furthermore, 
the synchrotron emission arises from a more
compact and presumably more homogeneous region.  The source may also
have less synchrotron self-absorption at millimeter wavelengths,
although this is not required by all models.
We will report
in a future paper on millimeter polarimetric observations of Sgr A*.

\acknowledgements 
We thank Alok Patnaik for enlightening discussions.
HF is supported by DFG grant Fa 358/1-1\&2.

\newpage

\begin{deluxetable}{llrrrr}
\tablecaption{Polarized and Total Flux from Continuum Observations 
at 4.8 GHz \label{tab:vla6cm}}
\tablehead{
\colhead{Source}  & \colhead{Date} & \colhead{I} & \colhead{P} & \colhead{p} & \colhead{$\chi$} \\
  &                & \colhead{(Jy)} & \colhead{(mJy)} & \colhead{(\%)} & \colhead{(deg)} \\
}
\startdata
Sgr~A*   & 10 Apr 98 & 0.49  &  $<0.63$ & $<0.13$ & \dots \\
	 & 18 Apr 98 & 0.47  &  $<0.34$ & $<0.08$ & \dots \\
1741-038 & 10 Apr 98 & 4.90  & 10.6  & 0.22 &  27 \\
	 & 18 Apr 98 & 4.94  & 15.3  & 0.31 &  24 \\
1748-253 & 10 Apr 98 & 0.48  &  8.4  & 1.8  &  -15 \\
         & 18 Apr 98 & 0.48  & 11.4  & 2.4  & -16  \\
GC441    & 10 Apr 98 & 0.044 &  $<0.09$ & $<0.20$ &  \dots \\
         & 18 Apr 98 & 0.043 &  $<0.12$ & $<0.28$ &  \dots  \\
W56      & 10 Apr 98 & 0.104 &  2.2  & 2.1 & -66 \\
	 & 18 Apr 98 & 0.104 &  2.2  & 2.1 & -66  \\
W109     & 10 Apr 98 & 0.098 &  0.59 & 0.6 & -26 \\
	 & 18 Apr 98 & 0.098 &  0.51 & 0.5 & -17  \\
\enddata
\end{deluxetable}

\begin{deluxetable}{lrrrr}
\tablecaption{Polarized Flux and Rotation Measure from Spectro-Polarimetric
Observations at 4.8 GHz \label{tab:rm6cm}}
\tablehead{
\colhead{Source}  &  \colhead{P} & \colhead{p} & \colhead{RM} & \colhead{$\sigma_{\rm RM}$} \\
                  & \colhead{(mJy)} & \colhead{(\%)} & \colhead{($\rdm$)} & \colhead{($\rdm$)} 
}
\startdata
3C 286 & $727 \pm 10$ & $11.2 \pm 0.2$ & $-7.3 \times 10^2$ & $2.7 \times 10^2$ \\
NRAO 530 & $106 \pm 1$ & $2.4 \pm 0.03$ & $-11.9 \times 10^2$ & $2.2 \times 10^2$ \\
Sgr A* & $<0.8$ & $<0.15$ & \dots  & \dots \\
\enddata
\end{deluxetable}

\begin{deluxetable}{lrrrr}
\tablecaption{Polarized Flux and Rotation Measure from Spectro-Polarimetric
Observations at 8.4 GHz \label{tab:rm3cm}}
\tablehead{
\colhead{Source}  &  \colhead{P} & \colhead{p} & \colhead{RM} & \colhead{$\sigma_{\rm RM}$} \\
                  & \colhead{(mJy)} & \colhead{(\%)} & \colhead{($\rdm$)} & \colhead{($\rdm$)} 
}
\startdata
3C 286 & $592 \pm 3$ & $12.1 \pm 0.06$ & $4.5 \times 10^2$ & $7.6 \times 10^2$\\
NRAO 530 & $101 \pm 1$ & $2.0 \pm 0.02$ & $-3.3 \times 10^3$ & $1.2 \times 10^3$\\
Sgr A* & $< 1.0$ & $<0.17$ & \dots & \dots  \\
W56 & $2.7 \pm 0.1$ & $2.9 \pm 0.1$ & $-8.7 \times 10^4$ & $12.9 \times 10^4$ \\
1741-312 & $20 \pm 2$ & $4.7 \pm 0.5$ & $6.1 \times 10^4$ & $9.0 \times 10^4$ \\
GC 441 & $<1.1$ & $<5.8$ & \dots  & \dots\\
W109 & $1.0 \pm 0.1$ & $1.1 \pm 0.1$ & $4.5 \times 10^5$ & $2.4 \times 10^5$ \\
1748-253 & $8.4 \pm 0.4$ & $3.1 \pm 0.1$ & $9.0 \times 10^4$ & $10.3 \times 10^4$ \\
\enddata
\end{deluxetable}

\end{document}